%% file: main.tex
\documentclass[conference]{IEEEtran}
\IEEEoverridecommandlockouts
\usepackage{cite}
\usepackage{amsmath,amssymb,amsfonts}
\usepackage{algorithmic}
\usepackage{graphicx}
\usepackage{caption}
\usepackage{subcaption}

\usepackage{tikz}
\usetikzlibrary{shapes.geometric, arrows, positioning}

\usepackage{xargs}
\pgfdeclarelayer{background}
\pgfdeclarelayer{foreground}
\pgfsetlayers{background,main,foreground}

\tikzstyle{startstop} = [rectangle, rounded corners, minimum width=2cm, minimum height=0.7cm,text centered, text width=3cm, draw=black]
\tikzstyle{io} = [trapezium, fill=white, trapezium left angle=70, trapezium right angle=110, minimum width=2cm, text width=2cm, inner ysep=5pt, text centered, draw=black]
\tikzstyle{process} = [rectangle, fill=white, minimum width=2cm, minimum height=0.7cm, text centered, draw=black, text width=3cm]
\tikzstyle{decision} = [diamond, minimum width=2cm, minimum height=0.7cm, text centered, draw=black,  text width=3cm]
\tikzstyle{data} = [cylinder, minimum width=2cm, minimum height=0.7cm, text centered, draw=black, fill=white, text width=3cm, shape border rotate=90, aspect=0.1]
\tikzstyle{arrow} = [thick,->,>=stealth]

\tikzstyle{materia}=[draw, fill=white, text width=6.0em, text centered,
  minimum height=1.5em,drop shadow]
\tikzstyle{etape} = [materia, text width=8em, minimum width=10em,
  minimum height=3em, rounded corners, drop shadow]
\tikzstyle{texto} = [above, text width=6em, text centered]
\tikzstyle{linepart} = [draw, thick, color=black!50, -latex', dashed]
\tikzstyle{line} = [draw, thick, color=black!50, -latex']
\tikzstyle{ur}=[draw, text centered, minimum height=0.01em]

\newcommand{\data}[2]{node (p#1) [data]
  {#2}}

\newcommand{\process}[2]{node (p#1) [process]
  {#2}}

\newcommand{\background}[5]{%
  \begin{pgfonlayer}{background}
    \path (#1.west |- #2.north)+(-0.5,0.25) node (a1) {};
    \path (#3.east |- #4.south)+(+0.5,-0.25) node (a2) {};
    \path[fill=black!10,rounded corners, draw=black!50, dashed]
      (a1) rectangle (a2);
      \path (#3.east |- #2.north)+(0,0.25)--(#1.west |- #2.north) node[midway] (#5-n) {};
      \path (#3.east |- #2.south)+(0,-0.35)--(#1.west |- #2.south) node[midway] (#5-s) {};
      \path (#3.east |- #2.north)+(0.7,0)--(#3.east |- #4.south) node[midway] (#5-w) {};
  \end{pgfonlayer}}

\usepackage{textcomp}
\usepackage{xcolor}
\def\BibTeX{{\rm B\kern-.05em{\sc i\kern-.025em b}\kern-.08em
    T\kern-.1667em\lower.7ex\hbox{E}\kern-.125emX}}

\begin{document}

\title{Reliving the Dataset: Combining the Visualization of Road Users’ Interactions with Scenario Reconstruction in Virtual Reality\\
\thanks{This publication was written in the framework of European Union’s Horizon 2020 Research and Innovation Programme under grant agreement no 815001, project Drive2theFuture (\emph{Needs, wants and behavior of "\textbf{Drive}rs" and automated vehicles users today and in\textbf{to the future}}) and also partially supported by the Intel Collaborative Research Institute for Safe Automated Vehicles (ICRI-SAVe).}
}

\author{\IEEEauthorblockN{Lars Töttel\IEEEauthorrefmark{1},
Maximilian Zipfl, Daniel Bogdoll,
Marc René Zofka, and J. Marius Zöllner}
\IEEEauthorblockA{Technical Cognitive Systems,
FZI Research Center for Information Technology\\
Karlsruhe, Germany\\
Email: \{toettel, zipfl, bogdoll, zofka, zoellner\}@fzi.de}}

\maketitle

\begin{abstract}
One core challenge in the development of automated vehicles is their capability to deal with a multitude of complex traffic scenarios with many, hard to predict traffic participants. As part of the iterative development process, it is necessary to detect critical scenarios and generate knowledge from them to improve the highly automated driving (HAD) function. In order to tackle this challenge, numerous datasets have been released in the past years, which act as the basis for the development and testing of such algorithms. Nevertheless, the remaining challenges are to find relevant scenes, such as safety-critical corner cases, in these datasets and to understand them completely.

Therefore, this paper presents a methodology to process and analyze naturalistic motion datasets in two ways:
on the one hand, our approach maps scenes of the datasets to a generic semantic scene graph which allows for a high-level and objective analysis. Here, arbitrary criticality measures, e.g. TTC, RSS or SFF, can be set to automatically detect critical scenarios between traffic participants.
On the other hand, the scenarios are recreated in a realistic virtual reality (VR) environment, which allows for a subjective close-up analysis from multiple, interactive perspectives.
\end{abstract}


\section{Introduction}
\label{sec:introduction}

\begin{figure}
  \centering
  \includegraphics[width=\linewidth]{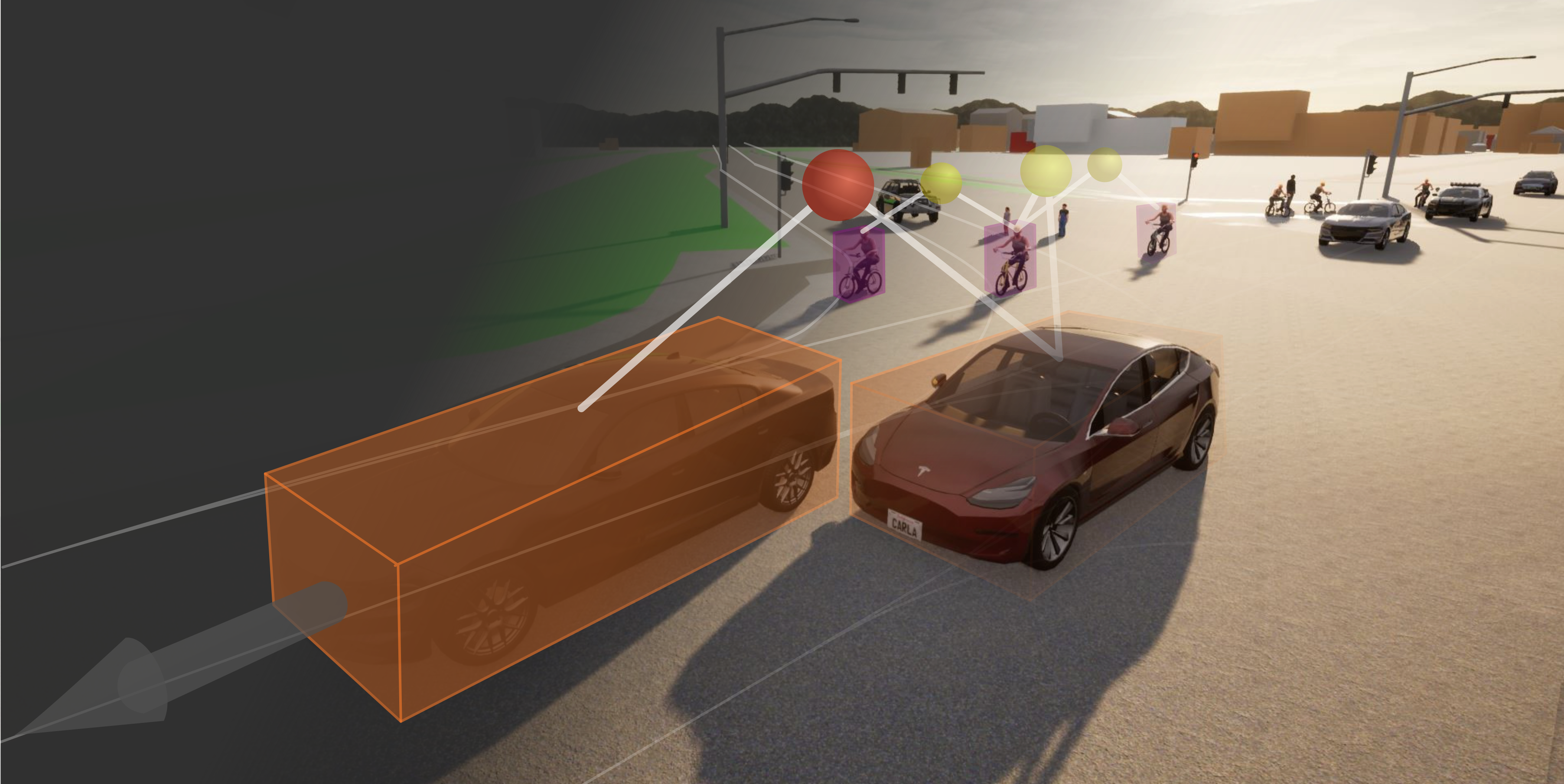}
  \caption{A critical scene that was first identified from a naturalistic motion dataset and later reconstructed in our simulation environment. The reconstruction allows for scenario exploration in virtual reality, where the user has full control over the time dimension and can relive the scenario from all possible points of view.}
  \label{fig:teaser}
\end{figure}

Naturalistic motion datasets allow for the analysis of behavior and interactions of traffic participants in a multitude of scenarios\footnote{We follow the definitions of scene and scenario by Ulbrich et al. \cite{Ulbrich2015}} and geographical areas. The data is mostly collected from infrastructure, drones or ego vehicles and sometimes combined with underlying high definition (HD) maps \cite{huang2020autonomous}. A general issue in automated driving related datasets is the heavy-tailed distribution of unusual, critical or surprising events. This makes knowledge discovery especially hard, since the detection and handling of such situations is key for the robustness of an autonomous system \cite{Koopman2018}. Critical situations are particularly relevant, as in such cases the safety of road users may be at risk. 

Besides finding critical situations such as corner-cases\footnote{We follow the conceptualization of corner-cases by Heidecker et al. \cite{heidecker2021applicationdriven}} within a dataset, a methodology is needed for the visualization of the scene considering all traffic participants as well as their interactions in an effective way. Thus, this paper proposes an approach to utilize a three-dimensional, spatio-temporal visualization based on the concept of \textsl{Space Time Cubes} \cite{Haegerstrand:1970, Kraak2003} for application in microscopic road traffic scenarios\cite{Junietz2019}. As a result, our approach enables and simplifies visual relation analysis of interacting traffic participants at street level. 

At the same time, we implement an interface to a high-fidelity simulator in order to recreate scenes from datasets in a realistic environment. Using virtual reality (VR), we enable users to immerse into the scene from any perspective, including pre-defined options such as the position of a driver of any vehicle or pedestrian included in the dataset. As a result, this integration allows for another close-up level of dataset analysis. In order to demonstrate this analysis approach, we also give an application example of a critical scene of a chosen dataset. 
To the best of the authors' knowledge, our approach is the first of its kind which makes use of VR in order to reconstruct automated driving related datasets for the purpose of close-up scenario analysis from multiple perspectives.

This paper is structured as follows: In Sec. \ref{sec:related_work}, we give an overview of the related work in the field of automated driving related datasets and their analysis. In Sec. \ref{sec:concept}, we introduce our concept for the analysis of road traffic datasets using various criticality measures. In addition, we propose an approach to visualize the analysis' results within a virtual reality environment. Afterwards, we show the realization within a simulation framework in Sec. \ref{sec:realization}, followed by an evaluation by applying our approach to an exemplary dataset in Sec. \ref{sec:evaluation}. Lastly, we give a summary and outlook in Sec. \ref{sec:conclusion}.

\section{Related Work}
\label{sec:related_work}
\subsection{Automated Driving Related Datasets}
\label{sec:related_work:data}
Automated driving related datasets are becoming more numerous, larger, diverse and have significantly more annotated frames \cite{Laflamme2019}. While the popular \emph{KITTI} dataset from 2012 consists of 22 scenes with 15.000 annotated frames, the more recent \emph{nuScenes} and \emph{Waymo Open} datasets consist of thousands of scenes with hundreds of thousands of annotated frames. The crowdsourced \emph{BDD100K} dataset even consists of 100.000 scenes \cite{Geiger2012, Sun2020, Caesar2020, Yu2020}. 

\subsection{Dataset Analysis}
\label{sec:related_work:data_analysis}
Two core challenges need to be addressed - how to detect data points, such as corner-cases, that are relevant for inspection, which are clearly the minority of the points and how to visualize complex data in a way humans can comprehend. A pipeline for the detection of relevant data in combination with a VR based visualization has the potential of addressing this issue \cite{VanDam2002}. The analysis of traffic-related datasets is split into two categories. The average occurrence of accidents or other risks is described on the macroscopic layer, while the risk of individual scenes are described on the microscopic layer \cite{Junietz2019}.

\subsection{Criticality Measures}
\label{sec:related_work:data_metrics}
Since we are interested in the detection of relevant scenes, the applied measures correspond to microscopic risk analysis. There are many classes of criticality measures, such as Time-To-X or sampling methods, with a variety of concrete implementations and combinations \cite{Junietz2019}. While these criticality measures are mainly utilized to identify critical situations, so called safety scores are designed to ensure the safety of vehicle states \cite{Zhao2020}. Both are well suited for a post-processing based analysis of datasets. These scores are based on safety policies, such as the \emph{Mobileye Responsibility-Sensitive Safety (RSS)} \cite{Shalev-Shwartz2017}, \emph{Nvidia Safety Force Field (SFF)} \cite{Nister2019} or \emph{TUM legal safety} frameworks \cite{Pek2020}. The various measures and scores vary in their ability to detect critical scenarios \cite{Junietz2019}, and thus can be dynamically exchanged or combined for analysis within our proposed framework.

\subsection{Dataset VR Exploration}
\label{sec:related_work:data_vr}

Virtual reality has a long history with many specialized forms and applications, starting in the 1960s \cite{Okechukwu2011}. More recently, mature and low cost virtual reality solutions have attracted the interest of the research community, leading to a growing number of VR-based applications \cite{Cipresso2018}. The visualization and analysis of high complexity datasets in VR can provide users with new insights and perspectives about the data and has been proven to be more successful and satisfying, compared to classical visualization methods \cite{Millais2018, Reski2020}. In the specific context of traffic scenarios and automated vehicles, virtual reality can be used for stated preference experiments \cite{Farooq2018}, which have been known to suffer from a lack of realism. It can also be used for pedestrian behavior studies, especially regarding risky situations \cite{Luo2020, Meir2015}. While some methods are able to import real road networks and generate virtual environments based on them, they lack the ability to utilize real recordings of vehicle trajectories \cite{Farooq2018}. 
To the best of our knowledge, there is only one VR application that fully imports automated driving related datasets for the specific purpose of improved data-labeling \cite{Wirth2019}. However, VR has not been used in order to explore motion datasets in a high-fidelity simulation environment.

\section{Concept}
\label{sec:concept}

\subsection{Analysis Framework}
\label{sec:concept:framework}

The analysis framework presented here is based upon naturalistic motion datasets. These mostly consist of object lists that describe traffic scenarios recorded over discrete timestamps. In addition, many of the datasets also provide high definition maps describing the road topology of the considered location.

Our proposed framework for visual analysis consists of two core components: an objective analysis (OA) module for the detection of critical scenes and a subjective analysis (SA) module for scenario recreation and knowledge discovery in VR. The resulting processing pipeline for naturalistic motion datasets is shown in Fig. \ref{fig:overview_framework}.

\input{figures/overview_tikz} 

For the objective analysis, we introduce a module that assesses traffic scenes by means of criticality measures and safety models to detect critical situations. Input data containing road users' trajectories is processed and abstracted. We implement an agnostic database interface which can deal with different kinds of input and datasets. Then, for each scene (i.e. discrete timestamp), a semantic scene graph is created, which is used to apply a criticality analysis based on various common measures. Finally, an abstract visualization of road users provides a comprehensive overview of the interactions between all traffic participants. The visualization consists of a 3D visualization of the scene graph's relations and the spatio-temporal visualization regarding traffic participants' trajectories.

For the subjective analysis, based on a transfer of the data into the virtual reality environment, particularly critical situations can be analyzed interactively and additional conclusions can be drawn. 
For example, a vehicle can occlude the driver's field of view, which might not have been obvious from an abstract, top-down perspective of the scene. Regarding this aspect, we make use of our existing co-simulation approach \cite{Zofka:2020} which is based on the Robot Operating System (ROS) \cite{Quigley:2009}. 

In the following sections, we introduce and describe all components of our OA and SA modules which are shown in Fig. \ref{fig:overview_framework}.

\subsection{Semantic Scene Graph}
\label{sec:concept:scene_graph}
As the first component of the OA module, our state description \cite{Zipfl:2020} is designed to be particularly efficient in determining the criticality of individual road users in relation to other road users. This generic abstraction allows it to apply almost freely exchangeable severity measures.

Objects of the given motion dataset are mapped to the lanes of the underlying road network. This leads to a reduction of the Cartesian space onto pseudo Frenet coordinates. On basis of the road topology given by the HD map, traffic participants can be brought into mutual relation. 

Relation types are being divided into three classifications: Vehicles traveling in the same lane (longitudinal relation), vehicles traveling in adjacent lanes (lateral relation), and vehicles travelling in intersecting lanes (intersection relation).

This state of a scene can then be described using a directed graph, where road users are described by nodes and their relationships by edges. A visualization of the graph can be seen in Fig. \ref{fig:3d_vis_scenegraph}. For more information regarding the composition of the semantic scene graph, the reader is referred to \cite{Zipfl:2020}. 

\subsection{Criticality Measures}
\label{sec:concept:metrics}
To enable safe automated driving, there is a need for robust perception and planning modules to handle a wide variety of situations. The development and training of such methods needs to incorporate corner case situations, which deviate from regular traffic scenes and are hard to find in large datasets. The most challenging corner cases for automated vehicles, according to \cite{heidecker2021applicationdriven}, are within the scene and scenario level. To detect relevant scenes within the scene graph, it is necessary to analyze the criticality of relations between traffic participants. Detected scenes can be used as a baseline for further evaluation of critical scenarios.

As the second component of the OA module, we present three different measures which are applicable within our approach: Time-To-Collision (TTC) \cite{iso:22839}, Responsibility-Sensitive Safety (RSS) \cite{Shalev-Shwartz2017} and Safety Force Field (SFF) \cite{Nister2019}. Our framework allows for the integration and combination of various measures and scores, since they vary in their ability to detect critical scenarios \cite{Junietz2019}. Therefore, this work focuses on the demonstration of analysis and visualization capabilities, instead of performing a full evaluation of the analysis results.

\subsubsection{Time-To-Collision}

Time-to-Collision (TTC) is one of the most commonly used criticality measures, which describes the time between two objects until they would collide with each other if they were to continue moving based on their current state \cite{Zheng2014}. In our case, we apply the constant velocity-$\textrm{TTC}_{A,B}$ between two objects ($A,B$), which is defined as follows:

\begin{align}
    \textrm{TTC}_{A,B} = \frac{d_{\overline{AB}}}{v_B-v_A}
\end{align}
where $d_{\overline{AB}}$ is the gap between object $A$ and object $B$ along the lane and $v_{A,B}$ describes their current velocity, respectively. 

TTC is usually applied for vehicles on the same road. However, the euclidean distance of two road users, who are not driving behind each other on the same lane, is mostly not sufficient to calculate the criticality through TTC. Therefore, a potential intersection point $p_i$ is created by predicting the road users according to their current driving direction.
Then, $p_i$ is used as a point in space for the calculation of the extension $\textrm{TTC}_{int}$ to $TTC$:

\begin{align}
    \textrm{TTC}_{int} = \frac{d_{\overline{B,p_i}}}{v_B}
    \label{eq:ttc_int}
\end{align}

Here, $v_B$ is the velocity of the vehicle that arrives second at the intersection point $p_i$ and the distance between the second arriving object $B$ and $p_i$ is denoted as $d_{\overline{B,p_i}}$ .

Equation \ref{eq:ttc_int} only applies when the condition, that both vehicles arrive at $p_i$ at the same time frame, is met.

Due to the discontinuity of the TTC function, the TTC value of an automated evaluation is often inappropriate if both velocities of the considered vehicle are identical $(v_A-v_B = 0)$. Therefore, the inverse Time-to-Collision $\textrm{TTC}^{-1}$ was presented in \cite{Balas2007}. The higher its value, the higher is the potential risk to cause an accident. $\textrm{TTC}^{-1}$ therefore provides a direct and continuous function for describing the collision risk.
Due to this advantage, we rather apply the inverse Time-to-Collision over its basic form.

\subsubsection{Responsibility-Sensitive Safety}
RSS \cite{Shalev-Shwartz2017} is a rigorous mathematical model that formalizes an interpretation of the law. RSS uses five common sense rules as a foundation for its formulization. In order to implement these, RSS establishes minimum safe distances and proper responses for different circumstances. Similar to TTC, RSS also calculates the safety distances based on the states of two road users. One special feature of the RSS model is that it calculates response times and acceleration values for individual road users.
In our approach, these vehicle parameters or person related values are assumed to be constant and identical for all road users of a class (\emph{Car}, \emph{Bike}, ...), since they are not included in the examined datasets.

In our implementation, the RSS model does not provide a continuous function as a result, but rather represents a binary state. As soon as a safety margin is fallen short of, the traffic constellation is classified as critical.

\subsubsection{Safety Force Field}

SFF \cite{Nister2019} is a safety-related theory to prevent collisions by obstacle avoidance, if applied. To utilize the SFF framework for the analysis of datasets, the \emph{safety potential} measure, which determines if actors are in a safe state, is used. 

In order to compute the \emph{safety potential} $\rho_{\rm A,B}$ between two actors $A$ and $B$, their safety procedure trajectories are checked for collisions and evaluated, see Eq. \eqref{eq:sff_computation}. Such a safety procedure can be modelled as a hard stop. The \emph{safety potential} between $A, B$ is 0, if no space-time collision is being detected. It is strictly positive in all other cases, based on the individual duration of the safety procedures $t_{\rm A_{\rm stop}}, t_{\rm B_{\rm stop}}$ and the time until they collide $c_{\rm t}$:

\begin{equation}
    \rho_{\rm A,B} = \left\Vert\left(t_{\rm A_{\rm stop}} - c_{\rm t}, t_{\rm B_{\rm stop}} - c_{\rm t}\rm\right) \right\Vert_{\rm p}
    \label{eq:sff_computation}
\end{equation}

While the complete SFF framework utilizes the measure and concept of it to avoid unsafe states, even for non-visible actors, the \emph{safety potential} measure is sufficient for the evaluation of scene criticality. 

Whereas a full implementation of the before mentioned RSS framework can be found publicly as free software \cite{Intel2021}, SFF is part of the commercial \emph{NVIDIA DriveWorks} framework \cite{Nvidia2020}. Therefore, we have implemented the \emph{safety potential} calculation ourselves as an exemplary case when both road users are driving on the same lane and thus have a longitudinal relation (see Sec. \ref{sec:concept:scene_graph}). The computation of the \emph{safety potential}, including the trajectories of the safety procedure \cite{Nister2019} and the collision checks \cite{Xu2011}, is fully implemented in the Frenet space. This allows us to implicitly utilize map-based lanes for the safety procedure.

\begin{figure*}[t]
\begin{subfigure}{.49\linewidth}
  \centering
  \includegraphics[width=\linewidth]{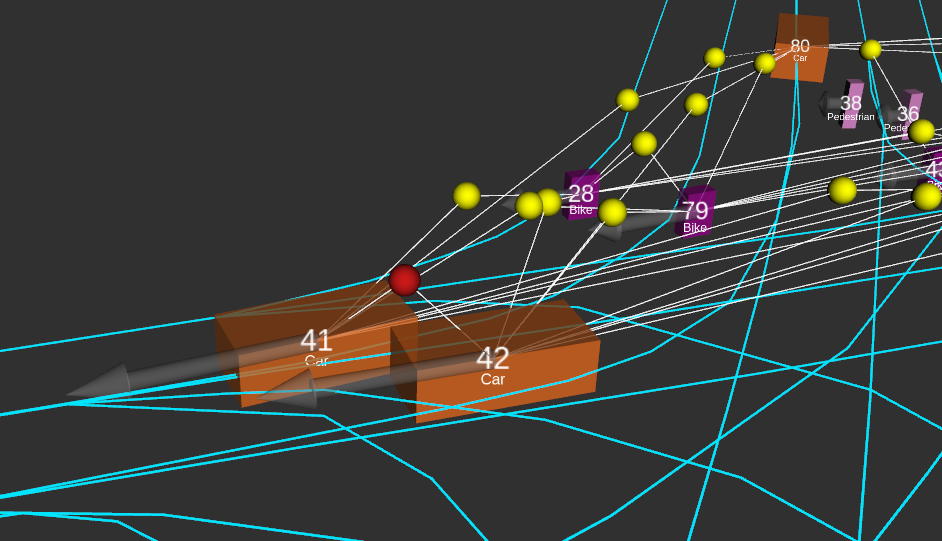}
  \caption{3D visualisation of a scene graph.\label{fig:3d_vis_scenegraph}}
\end{subfigure}
\hspace{\fill}
\begin{subfigure}{.49\linewidth}
  \centering
  \includegraphics[width=\linewidth]{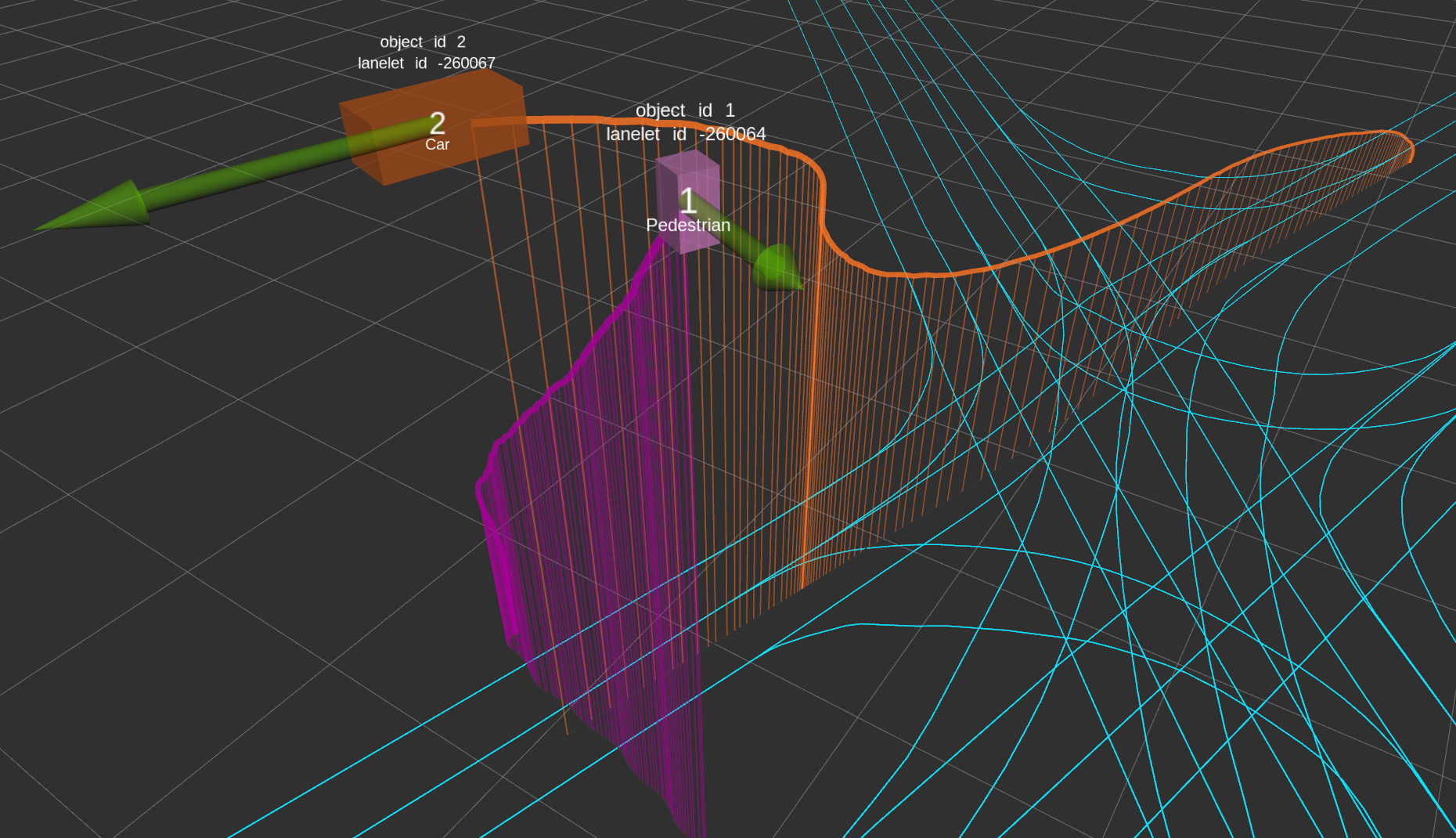}
  \caption{Spatio-temporal visualization of a scenario.\label{fig:spatio_temporal_vis}}
\end{subfigure}
\caption{Abstract visualizations from the OA module. In (a), the scene graph described in \ref{sec:concept:scene_graph} is being visualized. Traffic participants are displayed as colored boxes. Relations between them, built up by the scene graph, are displayed as white lines and colored spheres, which represent their criticality status. Darker spheres represent more critical relations. In (b), the spatio-temporal visualization of a scenario is shown. This enables a quick understanding of the scenario.\label{fig:abstract_vis}}
\end{figure*}

\subsection{Abstract Scene and Scenario Visualization}
\label{sec:concept:space_time_cubes}

As the third and final component of the OA module, we have implemented two abstract visualization functions for detected relevant scenes. First, we present the \emph{Semantic Scene Graph Visualization}, which displays the scene graph on top of the scene itself. This provides a fast overview of the relations between the road users. Second, we show our \emph{Spatio-Temporal Scenario Visualization}, which allows for fast insights into the surrounding scenario. Both types of visualization are shown in Fig. \ref{fig:abstract_vis}.

\subsubsection{Semantic Scene Graph Visualization}

The computed criticality between two traffic participants $A$ and $B$ are integrated into the visualization using a colored sphere between $A, B$ and additional edges to connect both traffic vehicles to the sphere. Here, darker spheres correspond to more critical relations, whereas lighter spheres represent less critical relations. Such a critical situation is shown in Fig. \ref{fig:3d_vis_scenegraph}.

\subsubsection{Spatio-Temporal Scenario Visualization}

For a spatio-temporal visualization for road traffic scenarios, the $z$ coordinate as well as the roll and pitch angles of traffic participants 
are neglected and only their two-dimensional $(x,y)$ position and yaw angle $\phi$ are used. Instead, the z-axis is used for the time dimension.

The visualization is done as follows. First, we visualize the road network on the spatial plane $(x,y,z=0)$. In order to better evaluate the spatial position of individual road users, the spatio-temporal trajectory is also projected onto the spatial plane $(x,y,z=0)$. Equal spatial positions of the spatio-temporal trajectory and the projected trajectory are connected by vertical lines. Inserting them every $n$ timesteps enables the viewer to get a better comprehension of the traffic participant's velocity: larger distances between two consecutive vertical lines represent higher velocities. For that reason, when a traffic participant only moves vertically upwards in the three-dimensional space, an accumulation of vertical lines at that point can easily be interpreted as a stop point as shown in Fig. \ref{fig:spatio_temporal_vis}.

\subsection{Scenario Recreation in a High-Fidelity Simulator}
\label{sec:realization:recreaton}

In order to visualize a dataset in a virtual reality environment, the complete scenario needs to be modelled. As the first part of the SA module, a methodology for its description is needed. First introduced in the \emph{Pegasus} research project \cite{Pegasus:2018}, Scholte et al. \cite{Scholtes:2021} refined the 6-layer model for the description of road traffic scenarios including the environment. Here, layer 1 represents the road network and layer 2 represents traffic infrastructure, such as traffic signs or trees next to the street. Layer 3 defines all temporary modifications of the first two layers, e.g. construction sites on the street. Level 4 describes all dynamic objects, level 5 describes the environment in terms of weather, lighting and other conditions. The final layer 6 is used for digital information, such as traffic light states or V2X information.  

\begin{figure}[t]
  \centering
  \includegraphics[width=0.49\textwidth]{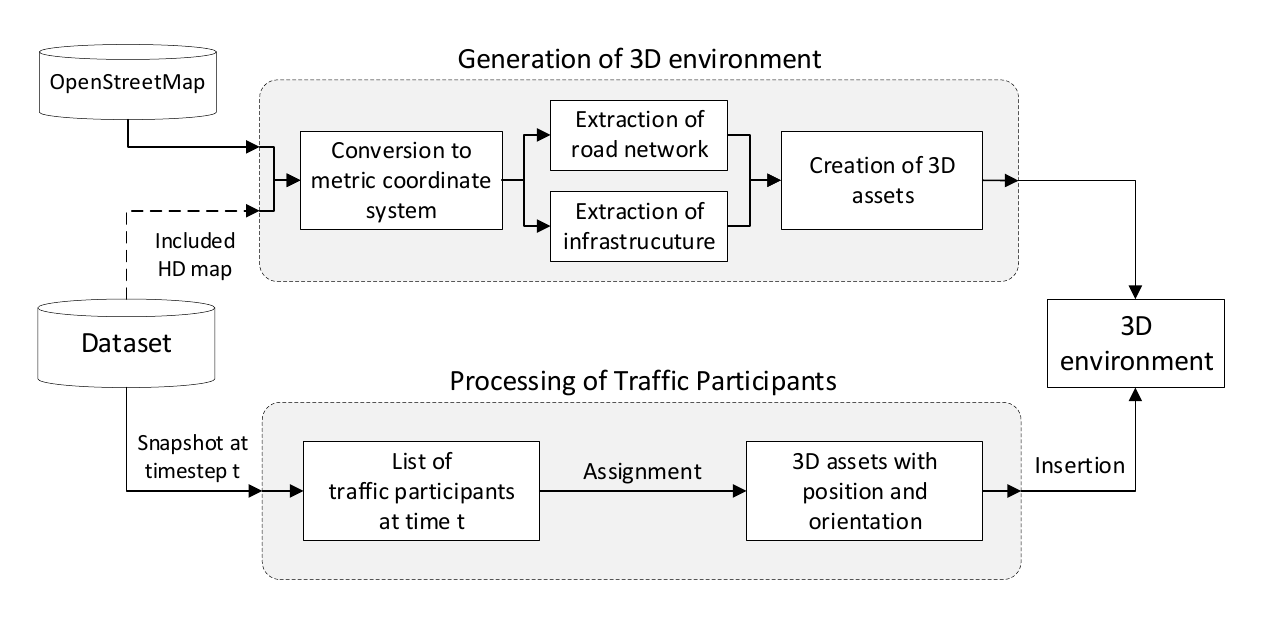}
  \caption{Workflow for the recreation of scenarios from datasets in a 3D simulation environment. Using the dataset and data from OpenStreetMap, both the static and dynamic environment can be reconstructed.}
  	\label{fig:workflow_recreation}
\end{figure}

Given a dataset, our approach to apply the above mentioned model is shown in Fig. \ref{fig:workflow_recreation}.
Here, multiple conditions must be fulfilled. For layer 1, 2 and 3, the dataset has to provide a high-definition (HD) map of the road network including relevant infrastructure in combination with a reference point in geodetic coordinates (latitude, longitude). 

Layer 4 concerns all traffic participants captured in the dataset. They must have at least a two-dimensional position and orientation for every timestep as well as a distinct classification, e.g. \emph{Car} or \emph{Pedestrian}. Consequently, the object type together with a position and orientation can be passed to the VR simulation in order to move a 3D model of the object in the simulated environment.\\
Weather or lighting described by layer 5 is almost never covered in naturalistic motion datasets, therefore, these effects can be controlled and varied completely by the VR simulation. If traffic light state information is available from the dataset, the data can be used for layer 6 and therefore be fed into the respective simulated traffic lights, otherwise we neglect the information and turn off all traffic lights.\\
An example of a dataset providing most of the required elements mentioned above, i.e. a high-definition map in \emph{Lanelet} \cite{Bender2014} format, a geodetic reference point, time-referenced object trajectories as well as traffic light information in the form of Signal Phase and Timing (SPaT) is the recently published TAF-BW dataset \cite{Zipfl:2020}.

\subsection{Knowledge Discovery in Virtual Reality}
\label{sec:realization:knowledge_vr}

As the second and final part of the SA module, we immerse into the reconstructed traffic scenarios using a virtual reality environment as shown in Fig. \ref{fig:picure_overview}. Here, we make use of the open-source simulator CARLA \cite{Dosovitskiy:2017} in version 0.9.11 which is widely used in automated driving research. CARLA provides assets for vehicles, cyclists and pedestrians, which we use to visualize all agents included in the input datasets. 

Furthermore, CARLA is based on Unreal Engine \cite{UnrealEngine} which allows us to make use of its VR capabilities. We implement a generic VR view in CARLA, where users can immerse into the simulation from the viewpoints of the inserted actors, i.e. as pedestrians or drivers.

In addition, using CARLA and Unreal Engine not only allows us to render the scene in a high-fidelity 3D environment and influence environmental conditions such as the weather, but also enables the simulation of sensor data which is not included in the dataset. For this, CARLA allows for the attachment of different sensors, e.g. cameras, radars or LIDARs, to any actor. Consequently, our approach also enables data augmentation for the training of machine learning based approaches, as it also can be used for sensor data generation. Furthermore, CARLA provides different weather conditions, enabling the exploration of scenarios from the dataset under different conditions.

\section{Realization}
\label{sec:realization}

\subsection{Objective Analysis Module}
\label{sec:realization:simulation_framework}
 
We realized the OA module, as shown in Fig. \ref{fig:dataset_analysis}, using ROS. Input from a naturalistic motion dataset is processed via a dataset-agnostic interface and converted to respective ROS types in order to allow handling different types of datasets. By using ROS, 
input data can also come directly from intelligent infrastructure, for example as implemented in the Test Area Autonomous Driving Baden-Württemberg (TAF-BW) \cite{Fleck:2018}. 

Based on this, the semantic scene graph is constructed and the criticality analysis is performed. The criticality analysis is done by continuously evaluating and integrating different criticality measures for the currently chosen timestamp. 
We have implemented the \emph{Semantic Scene Graph} and \emph{Spatio-Temporal Scenario} visualizations using RViz. 
For the semantic scene graph visualization, we buffer all timestamps from the input data and 
allow the user to interactively change the currently chosen value
in order to gain full control over the time dimension. A visualization of the semantic scene graph on top of the scene is done by means of colored spheres connected to two corresponding traffic participants: darker spheres symbolize more critical interactions, whereas lighter ones correspond to less critical interactions. The used criticality measure can be chosen.

Moreover, in order to not overload the visualization in the case of many traffic participants, we also add a threshhold parameter for the criticality measure to limit the number of spheres and edges included in the visualization. 
As a result, it is possible to loop through the input data and only visualize identified critical interactions, whereas all others can be ignored. Thereby, it is possible to precisely find and investigate critical timestamps. Evaluation results, such as the outputs of the applied measures and the corresponding timestamps, are saved for later reuse.

For the investigation of the entire scenario, i.e. of more than one timestamp, our spatio-temporal visualization is applied. This visualization includes past trajectories of all traffic participants, and eventual spatial conflicts can be resolved by using the time as the z-axis (see Fig. \ref{fig:spatio_temporal_vis}).

In summary, using our visual analysis framework, users can inspect all timestamps included in a dataset and inspect critical scenes using abstract, three-dimensional visualizations. 

\subsection{Subjective Analysis Module}
\label{sec:realization:subjective-analysis}

For the subjective analysis consisting of the recreation of scenarios and the immersion in virtual reality, we make use of HD maps and create 3D worlds within Unreal Engine.
Common open-source HD map data formats are \emph{OpenDRIVE} \cite{dupuis2010opendrive} or \emph{Lanelet} \cite{Bender2014}, which can in turn be parsed in order to create a model of the road. If no map is provided, a geodetic reference point or background knowledge about the location of the dataset can be used to create the road model from \emph{OpenStreetMap} data. However, this data is not as accurate as HD maps and might not cover all information. 

Afterwards, additional elements such as buildings or vegetation are added to the 3D world either manually or automatically by again parsing and converting data from OpenStreetMap. 
The 3D world is then integrated into CARLA, and traffic participants are now inserted and moved by exchanging ROS messages. Here, we make use of our ROS-based co-simulation architecture \cite{Zofka:2020}: CARLA is integrated as a visualization model, and an additional module is used to integrate traffic participants from the given dataset into the 3D world at any timestamp. 

\begin{figure}[t]
  \centering
  \includegraphics[width=\linewidth]{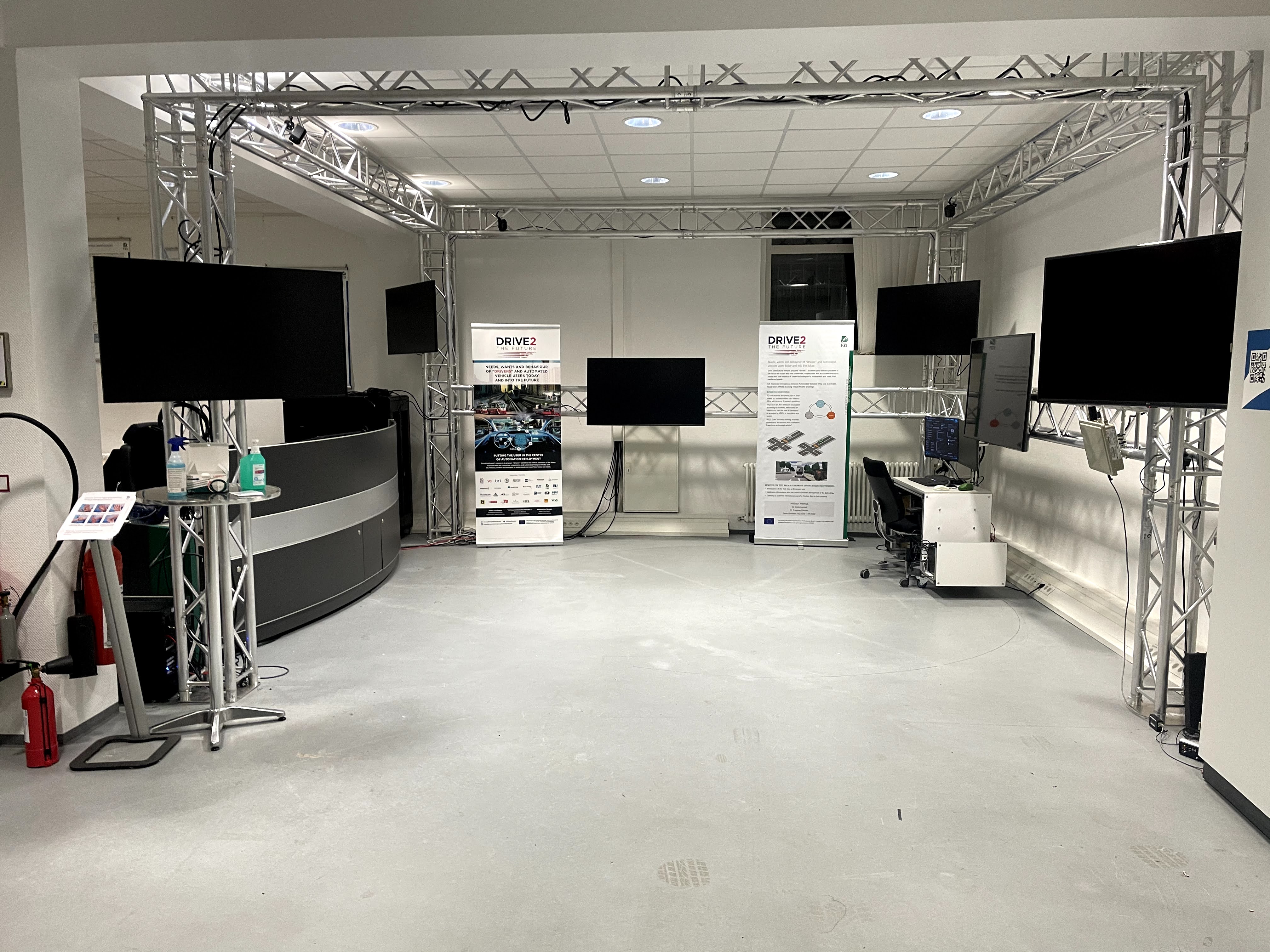}
  \caption{The virtual reality setup in our laboratory. The setup provides a walking area in order to move within the recreated traffic constellation.}
  \label{fig:vr_cave}
\end{figure}

For knowledge discovery in VR, we make use of our VR hardware setup introduced in \cite{Zofka:2020} and shown in Fig. \ref{fig:vr_cave}. We utilize a HTC VIVE VR headset and four SteamVR base stations, and data is transmitted wirelessly to the headset. The setup includes both a walking area for the immersion as a pedestrian or external observer and a driver's seat for the immersion as a driver.

Moreover, CARLA's architecture allows us to setup multiple, distributed computers for the VR environment: we run the simulation server including rendering on a computer running Windows 10 since many VR headset only support Windows, and we run the dataset processing and interaction analysis on another computer running Linux.   

\section{Evaluation}
\label{sec:evaluation}

For the demonstration of our methodology, we chose the following approach: first, we analyze a given dataset using the criticality measures described in Sec. \ref{sec:concept:metrics}. Second, we visualize one of the most critical timestamps we found. Third and last, we recreate the same setting in the VR environment in order to obtain an even deeper insight into the scene and the surrounding scenario.

\begin{figure}[t]
  \centering
  \includegraphics[width=\linewidth]{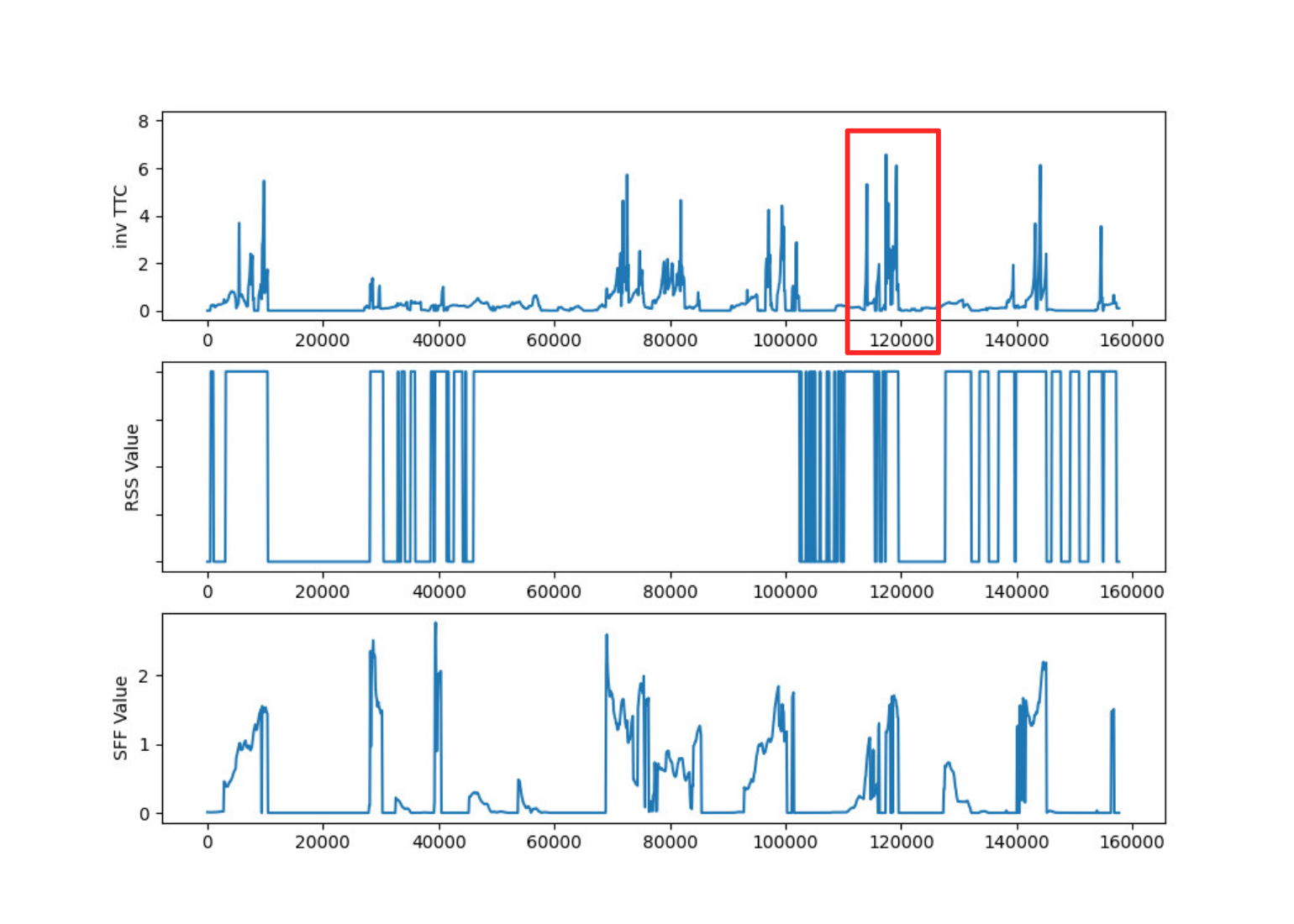}
  \caption{Criticality analysis of all timestamps of the TAF-BW dataset using the inverse TTC, RSS and SFF models. The red box highlights an exemplary critical interval.}
  \label{fig:dataset_analysis}
\end{figure}

The dataset we analyze using the criticality measures from Sec. \ref{sec:concept:metrics} is the Test Area Autonomous Driving Baden-Württemberg dataset \cite{Zipfl:2020}.
The results of our analysis using inverse TTC, RSS and SFF are depicted in Fig. \ref{fig:dataset_analysis}. 
Here, in all three diagrams, the x-axis shows the timestamps included in the dataset. The computed value of the respective measure is plotted on the y-axis, using the maximum value of all traffic participants for the respective timestamp. Both the inverse TTC and the SFF model describe more critical scenes with higher criticality values. Since the RSS model provides a binary output, each traffic scene in which any traffic participant falls below a safe-distance threshold is marked as critical and displayed as a high.

\begin{figure}[t]
  \centering
  \includegraphics[width=\linewidth]{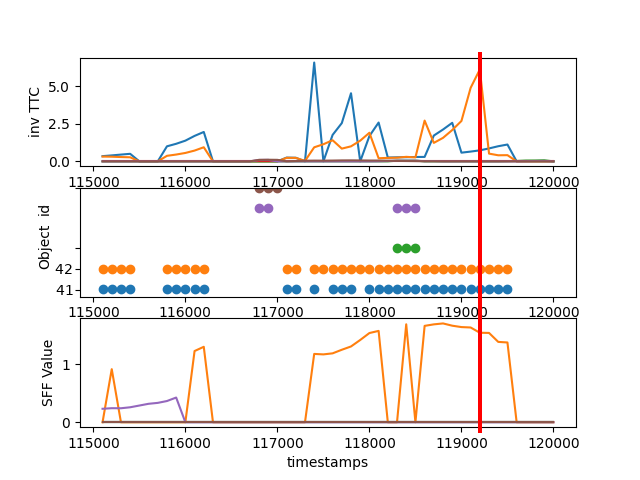}
  \caption{Close up analysis of the highlighted interval (see Fig. \ref{fig:dataset_analysis}). The criticality status (inverse TTC, RSS, SFF) of each individual traffic participant visible in this interval is plotted. We identify $t_2=119200~ms$ as the most critical timestamp according to TTC, which is highlighted by the  red line.}
  \label{fig:local_dataset_analysis}
\end{figure}

While TTC and SFF classify the criticality of scenes differently, significant overlaps can be seen in multiple timestamps and time intervals. 
However, there are also intervals that are only critical in one of the two measures, e.g. in $t \in [27000~ms, 43000~ms]$, there are two very critical scenes according to SFF, but they are much less critical according to TTC. 
Concurrently, RSS also yields peeks for all of these timestamps.

We chose to use the maximum value of the inverse TTC measure for our demonstration, and it can be seen that it is within $t_1=117400~ms$ and $t_2=119200~ms$, highlighted by a red box. 
Therefore, we investigate this time interval in more detail in Fig. \ref{fig:local_dataset_analysis}.  

Here, the courses of inverse TTC, RSS and SFF for all involved traffic participants are depicted. All measures identify the vehicles with $track\_id = 42$ (orange lines) and with $track\_id = 41$ (blue lines) as the ones interacting critically in the time interval. 
In the first peak of inverse TTC at $t_1=117400~ms$, the computed Time-To-Collision is $0.154~seconds$ and at the second peak at $t_2=119200~ms$, the TTC is $0.164~seconds$. Since the second peak is also identified by the SFF measure, we chose $t_2=119200~ms$ for our demonstration, which is the scene shown in Fig. \ref{fig:teaser}.

\begin{figure*}[ht]
\begin{subfigure}{.49\linewidth}
  \centering
  \includegraphics[width=\linewidth]{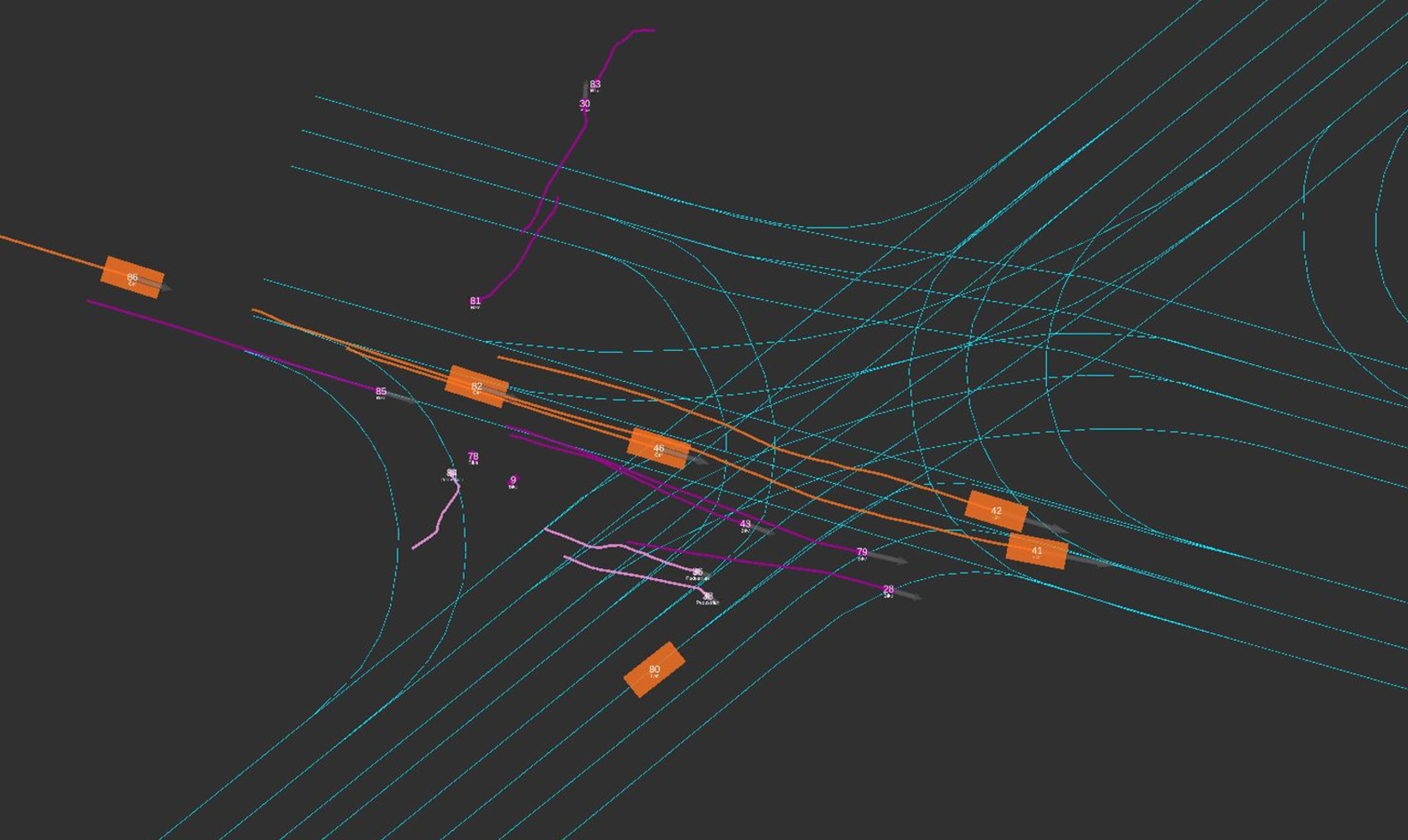}
  \caption{Top-down perspective of the abtract scene.\label{fig:eval:0}}
\end{subfigure}
\begin{subfigure}{.49\linewidth}
  \centering
  \includegraphics[width=\linewidth]{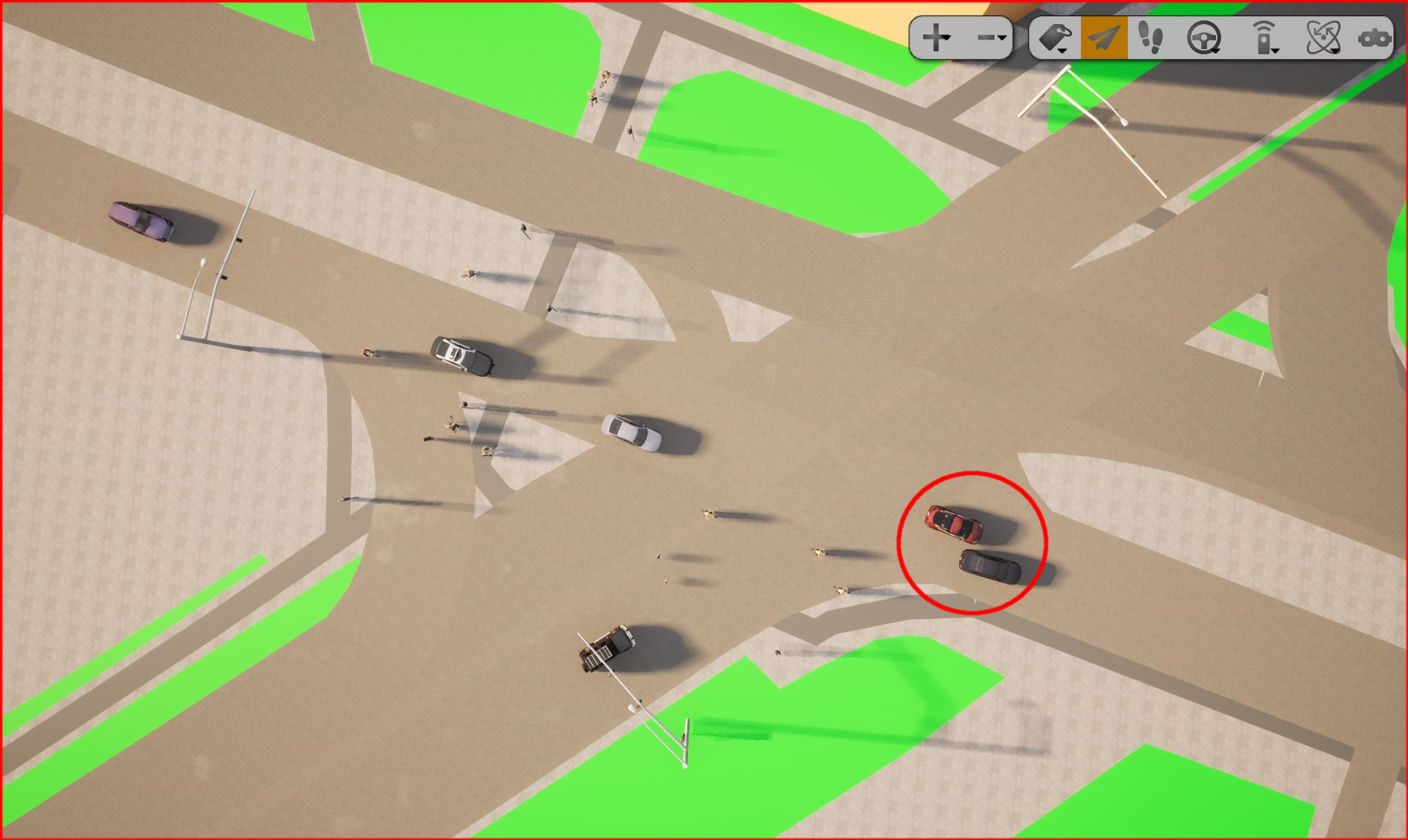}
  \caption{Top-down perspective of the reconstructed scene.\label{fig:eval:1}}
\end{subfigure}
\begin{subfigure}{.49\linewidth}
  \centering
  \includegraphics[width=\linewidth]{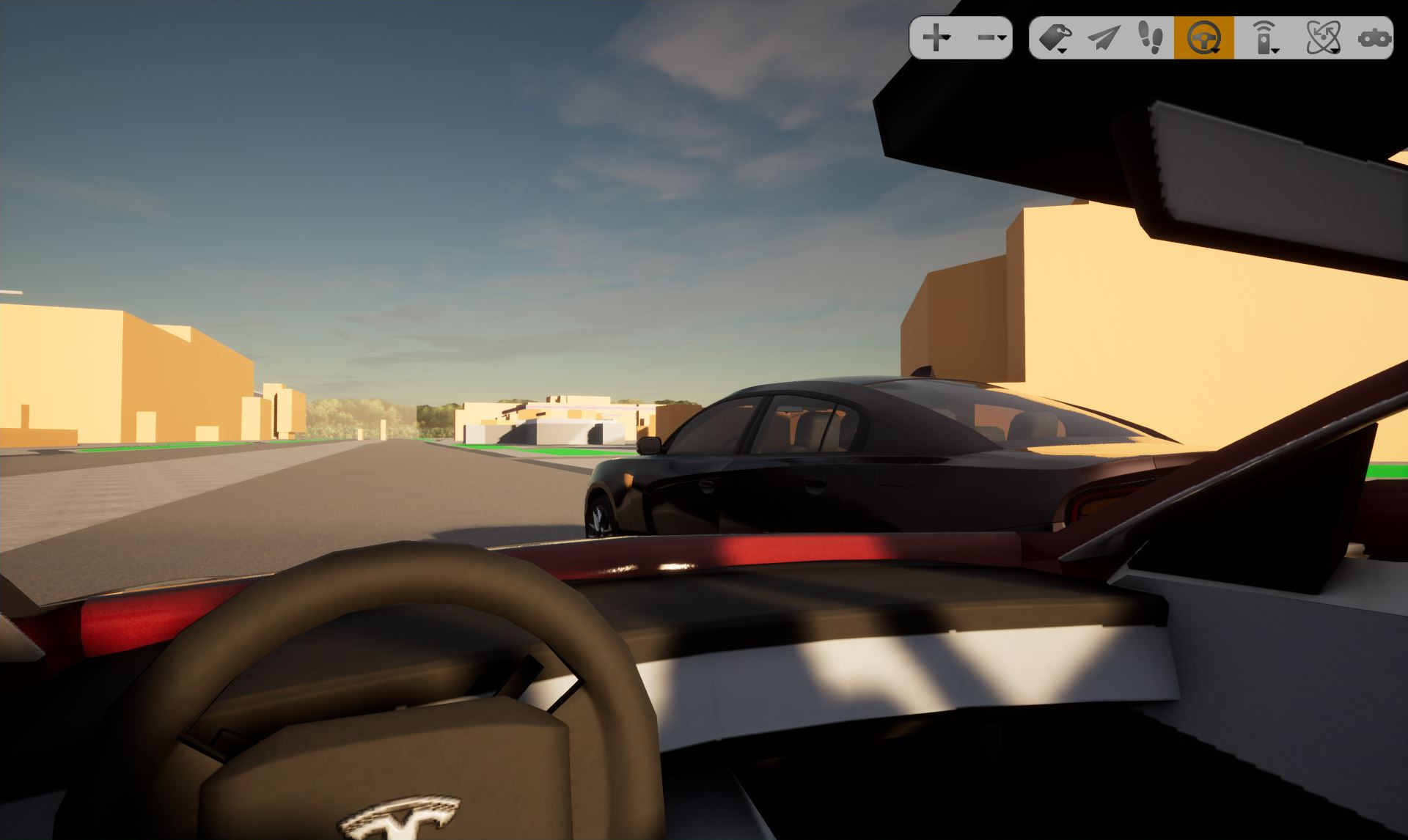}
  \caption{Front view of immersive driver of vehicle with $track\_id = 42$.\label{fig:eval:2}}
\end{subfigure}
\hspace{\fill}
\begin{subfigure}{.49\linewidth}
  \centering
  \includegraphics[width=\linewidth]{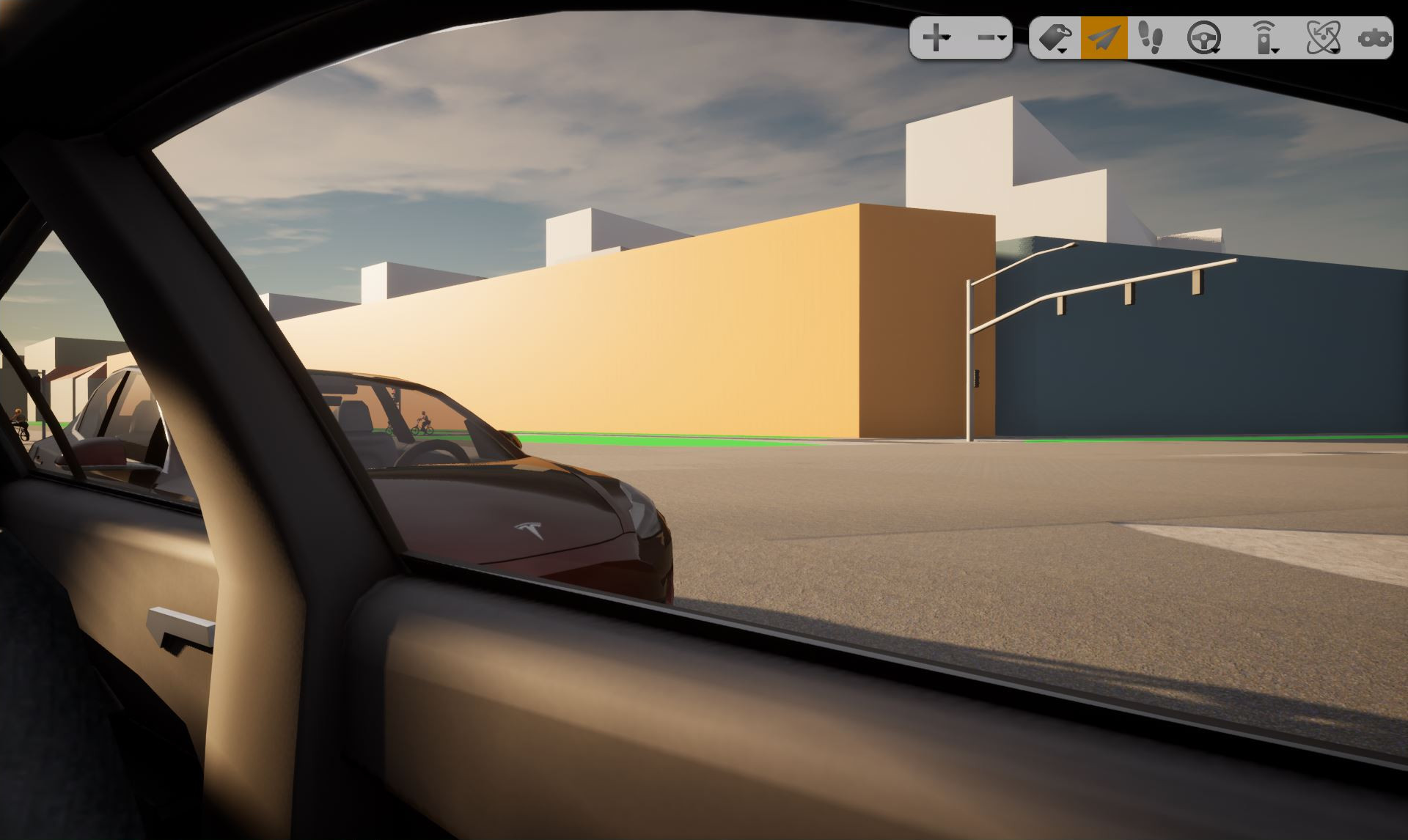}
  \caption{Side view of immersive driver of vehicle with $track\_id = 41$.\label{fig:eval:3}}
\end{subfigure}
\caption{The reconstructed scene from the TAF-BW dataset at $t_1=119200~ms$. In (a), an abstract top-down perspective of the scenario visualized in RViz is shown. In (b), the same scene is shown after it was reconstructed within our simulation environment. 
Here, the two critically interacting vehicles are highlighted by a red circle: the vehicle with $track\_id = 42$ (Tesla Model 3, red vehicle) moves in parallel and very close to the vehicle with $track\_id = 41$ (Dodge Charger, black vehicle). 
In (c) and (d), different fields of vision from the driver's seats of both vehicles are shown. The immersion in VR allows the user to look around and to closely investigate the traffic constellation. \label{fig:picure_overview}}
\end{figure*}

To visually analyze the critical scene at $t_2=119200~ms$, the objective analysis module is utilized. In Fig. \ref{fig:3d_vis_scenegraph}, a close up of the critical scene is shown. Relations built up by the scene graph are displayed as white lines connecting two traffic participants by means of a colored sphere (see Sec. \ref{sec:concept:scene_graph}). 
When a relation becomes more critical, the sphere turns from yellow to red (see relation between vehicle $id = 41$ and vehicle $id = 42$). The abstract top-down visualization in Fig. \ref{fig:eval:0} shows that the vehicles are driving in parallel and that they are very close to each other. 

Finally, we recreate the scenario in our simulation environment and immerse into the situation using a VR head-mounted display. Snapshots of the critical scene at $t_2=119200~ms$ are shown in Fig. \ref{fig:picure_overview}. Now, in the VR environment, the user is consecutively seated at the driver's seat of the two critically interacting vehicles, where he can look around and therefore is able to gain more subjective insights into the traffic scenario. In addition, the user also has full control of going backwards and forwards in time.

In our example, we conclude that the critical scene resulted from two vehicles driving in parallel and very close to each other. The effect was enforced due to the fact that the preceding vehicle was not driving in parallel to its own lane, but rather turned slightly to the left lane. One possible explanation of this critical situation could be that the red vehicle with $track\_id = 42$ was in a blind spot and could not be seen by the driver of the vehicle with $track\_id = 41$. However, the vehicles' positions and orientations might also have been affected by sensor noise.

In summary, the objective analysis revealed critical scenes as input for the visual analysis. The abstract visualization allows the user to generate knowledge about the scene and the criticality relations between the road users. The final reconstruction of the scene in simulation and VR enables a more subjective and deeper exploration of the scene from different perspectives. In particular, driver perspectives of the vehicles with high-criticality relations provide a high level of insight and are available as presets in the SA module.

\section{Conclusion and Outlook}
\label{sec:conclusion}

Within the objective analysis (OA) module, we have proposed a combination of the analysis of interactive driving scenarios from naturalistic motion datasets, using various criticality measures, with two types of visualization approaches. We applied multiple criticality measures in order to specifically select scenes within the dataset which are worth investigating. On the one hand, we used abstract visualization methods in order to inspect the relations and criticality between interacting traffic participants as the last part of the OA module. On the other hand, we also reconstructed the scene in a virtual reality environment in order to inspect the recorded traffic situation from the perspective of drivers and consequently to obtain subjective insights into the scene as final part of the subjective analysis (SA) module. 

The resulting approach can be applied to common naturalistic motion datasets such as INTERACTION \cite{Zhan:2019}, TAF-BW \cite{Zipfl:2020} or inD \cite{inDdataset}. 
Furthermore, it yields a comprehensive visualization of the interactions between traffic participants using arbitrary measures such as TTC, RSS and SFF, making it easier to find corner-cases, such as critical situations, and understand the roles of the respective traffic participants in the scene or scenario.
Due to the modular structure of our approach, some components can also be utilized for the analysis of online data streams.
This might be the case for processed sensor data recorded by an automated vehicle, but also for data coming from intelligent infrastructure. Such infrastructure is available, for example, in the Test Area Autonomous Driving Baden-W\"urttemberg \cite{Fleck:2018}. 

In future studies, we will apply the proposed approach in scenarios including both automated vehicles, non-automated vehicles and vulnerable road users. Since our implementations of TTC, RSS and SFF are only examples to demonstrate the functionality, a proper combination of analysis measures might be considered in the future. Consequently, building upon the knowledge discovered from our OA and SA modules, we plan to generate a realistic, virtual dataset containing such traffic scenarios with focus on vulnerable road users. This includes, amongst others, situations where pedestrians disobey traffic rules and/or traffic lights.

\bibliographystyle{IEEEtran}
\bibliography{references}
\vspace{12pt}

\end{document}

%% file: figures/overview_tikz.tex
\begin{figure}[htbp]
\begin{center}
\begin{tikzpicture}[scale=0.7,transform shape]

  
  \node (p1) {};
  \path (p1) + (-4,-1.23)\data{2}{Naturalistic\\motion dataset};

  \path (p1.south)+(0.0,-1.1) \process{4}{Semantic scene graph};
  \path (p4.south)+(0.0,-1.1) \process{5}{Criticality analysis};
  \path (p5.south)+(0.0,-1.1) \process{7}{Abstract visualization};

  \path (p7.south)+(0.0,-1.5) \process{8}{Scenario recreation in a high-fidelity simulator};
  \path (p8.south)+(0.0,-1.1) \process{9}{Knowledge\\discovery in VR};


  \path [line] (p7.east) -- + (1,0) 
                |- node[pos=0.25, right] {}  (p8);
                
  \path [line] (p2.east) -- + (0,0) 
                |- node[pos=0.25, right] {}  (p4);
                
  \path [line] (p2.south) -- + (0,0) 
                |- node[pos=0.25, right] {}  (p7);
                
  \path [line] (p2.south) -- + (0,0) 
                |- node[pos=0.25, right] {}  (p8);

  \background{p4}{p4}{p7}{p7}{bk1}
  \background{p8}{p8}{p9}{p9}{bk2}
  
  \node[right=0.3cm, align=center] at (bk1-w.east) {Objective Analysis:\\Detection of Critical Scenes};
  \node[right=0.3cm, align=center] at (bk2-w.east) {Subjective Analysis:\\Scenario Exploration in VR};

  \path [line] (p4.south) -- node [above] {} (p5);
  \path [line] (p5.south) -- node [above] {} (p7);
  
  \path [line] (p8.south) -- node [above] {} (p9);


  
\end{tikzpicture}
    \end{center}
\caption{Overview of the framework for visual analysis. Using naturalistic motion datasets, we apply both an objective and a subjective analysis. The objective analysis module is used to find critical scenes within the dataset using different metrics and provides abstract visualisation options for scenes and their surrounding scenarios. The subjective analysis module allows for a reconstruction and an in-depth investigation of these scenes in a VR environment.}
\label{fig:overview_framework}
\end{figure}